\documentclass{article}

\usepackage{arxiv}

\usepackage[utf8]{inputenc} 
\usepackage[T1]{fontenc}    
\usepackage{hyperref}       
\usepackage{url}            
\usepackage{booktabs}       
\usepackage{amsfonts}       
\usepackage{nicefrac}       
\usepackage{microtype}      
\usepackage{lipsum}
\usepackage{graphicx}
\graphicspath{ {./images/} }
\usepackage{natbib}
\usepackage{doi}
\graphicspath{ {./images/} }
\usepackage{float}
\usepackage{placeins}
\usepackage{dblfloatfix}

\title{Decoding BACnet Packets: A Large Language Model Approach for Packet Interpretation}

\author{
 {\hspace{1mm}Rashi Sharma}\\
	Panasonic R\&D Center
        Singapore\\
	\texttt{rashi.sharma@sg.panasonic.com} \\
	\And
	{\hspace{1mm}Hiroyuki Okada} \\
	Panasonic Holdings Corporation\\		\texttt{okada.hiroyuki001@jp.panasonic.com} \\
 \And
  {\hspace{1mm}Tatsumi Oba} \\
	Panasonic Holdings Corporation\\		\texttt{oba.tatsumi@jp.panasonic.com} 
  \And
  {\hspace{1mm}Karthikk Subramanian} \\
	Panasonic R\&D Center
        Singapore\\		\texttt{karthikk.subramanian@sg.panasonic.com} 
        \And
        {\hspace{1mm}Naoto Yanai} \\
	Panasonic Holdings Corporation\\
\texttt{yanai.naoto@jp.panasonic.com} \\
 \And
 {\hspace{1mm}Sugiri Pranata } \\
	Panasonic R\&D Center
        Singapore\\	\texttt{sugiri.pranata@sg.panasonic.com} \\
}

\begin{document}
\maketitle
\begin{abstract}
The Industrial Control System (ICS) environment encompasses a wide range of intricate communication protocols, posing substantial challenges for Security Operations Center (SOC) analysts tasked with monitoring, interpreting, and addressing network activities and security incidents. Conventional monitoring tools and techniques often struggle to provide a clear understanding of the nature and intent of ICS-specific communications. To enhance comprehension, we propose a software solution powered by a Large Language Model (LLM). This solution currently focused on BACnet protocol, processes a packet file data and extracts context by using a mapping database, and contemporary  context retrieval methods for Retrieval Augmented Generation (RAG). The processed packet information, combined with the extracted context, serves as input to the LLM, which generates a concise packet file summary for the user. The software delivers a clear, coherent, and easily understandable summary of network activities, enabling SOC analysts to better assess the current state of the control system.
\end{abstract}


\section{Introduction}
In industrial control systems (ICS), packets are vital for communication and control. Various protocols, such as Modbus, PROFINET, and BACnet, define how packets are structured and handled in ICS. These protocols ensure that devices from different manufacturers can work together. However, the diversity and complexity of these communication protocols present significant challenges for Security Operations Center (SOC) analysts. They struggle to monitor, understand, and respond to network activities and security incidents effectively. Traditional monitoring tools and methods make it hard to grasp the nature and intent of ICS-specific communications. This can lead to issues like difficulty understanding the current state of the control system from communication logs, which can result in failure to identify threats and respond promptly. Additionally, providing clear explanations to non-technical stakeholders can be challenging, potentially delaying decision-making. In this paper, we present a Large Language Model (LLM) based solution to improve the understanding of packet data.

Large Language Models (LLMs) are advanced AI systems designed to understand, generate, and interact with human language in a meaningful and contextually appropriate manner \citep{overviewLLM24}. These models are trained on vast amounts of text data using advanced machine learning techniques, such as semi-supervised and self-supervised learning. By learning to predict the next word or character in a sequence, LLMs can generate coherent and relevant responses to a wide range of prompts \citep{Openai}. LLMs have demonstrated remarkable proficiency in various Natural Language Processing (NLP) tasks, including translation, summarization, question answering, and text generation. They have the potential to perform a broad spectrum of tasks specified in natural language, from composing emails and writing essays to explaining complex concepts and engaging in dialogues.

There are multiple methods to use LLMs for targeted applications. One common method is fine-tuning the LLM(\citep{FinetuningLLM} provides a guide on finetuning for enterprise applications)  on a specific dataset to generate text that is relevant, coherent, and informative. However, fine-tuning an LLM can be computationally expensive, requiring significant resources in terms of hardware, time, and energy. This can make it difficult to fine-tune large models or experiment with different strategies. To avoid the computation costs of fine-tuning, we use Retrieval-Augmented Generation (RAG) \citep{RAGK}. The RAG approach combines the strengths of both retrieval-based and generative models, improving the accuracy and relevance of generated responses by augmenting the language model with a retrieval mechanism that can access external knowledge sources.

In this paper we address the problem of interpreting network packets, specifically BACnet\footnote{\href{https://www.ashrae.org/technical-resources/bookstore/bacnet}{BACnet : https://www.ashrae.org/technical-resources/bookstore/bacnet}} packets using RAG for LLM. For now we are limiting the problem to BACnet because our business requires monitoring at the BACnet protocol level. This gives us the foundation for evaluating the usefulness of the software. Thus, the solution in this paper focuses on providing summaries for BACnet packets. 

BACnet is a networking protocol developed by ASHRAE for building automation and control systems. BACnet is an ICS protocol, which is used for applications like heating, ventilation, air-conditioning control, lighting control, access control, and fire detection systems. The protocol enables information exchange between computerized devices, regardless of their specific functions 
within the building services. Thus, BACnet can be effectively utilized by head-end computers, versatile direct digital controllers, and specialized or unitary controllers alike.

A BACnet packet contains two key components:

\begin{itemize}
	\item \textbf{BACnet NPDU (Network Protocol Data Unit):} Contains network layer information, including the version, control information, source, and destination addresses.
	\item \textbf{BACnet APDU (Application Protocol Data Unit):} Contains application layer information, including the service request or response.	
\end{itemize}

We extract the information from these key components of the BACnet packet and use it for context extraction for RAG. The context, combined with packet data, is passed to an LLM to generate a packet explanation.

With this solution, we aim to provide clear, concise, and understandable summaries of network activities and anomalies, allowing SOC operators to quickly grasp the current state of the control system and identify potential security threats. This will contribute to rapid triage and incident response. The explanations can also serve as a valuable training tool for new analysts, helping them understand complex network behaviours and security incidents through clear and detailed explanations. Additionally, the natural language summaries can be easily incorporated into reports for stakeholders, providing clear and actionable insights without requiring deep technical knowledge.

\textbf{Paper Organization}. The rest of this paper is organized as follows: Background information on technologies used in the solution are provided in Section \ref{sec:background}, Related studies are described in Section \ref{sec:related work}. The problem statement is outlined in Section \ref{sec:problem statement}. Section \ref{sec:Proposed Solution} describes the proposed solution. Experiments and results are provided in \ref{sec:Experiments and Results}. Section \ref{sec:Conclusion} concludes this paper. Lastly a few examples of packet summary generated by the proposed solution are available in the Appendix \ref{sec:Examples}.

\section{Background}
\label{sec:background}

\subsection{Large Language Models}
Large language models (LLMs) have recently shown remarkable performance in various natural language processing (NLP) tasks. These models are trained on vast amounts of text data and can generate human-like text, answer questions, translate languages, and even write code. Models such as BERT \citep{Bert}, GPT-2 \citep{Openai}, and T5 \citep{T5} have demonstrated the ability to capture complex linguistic patterns, enabling them to generate human-like text and perform tasks such as machine translation, question answering, and sentiment analysis with high accuracy.

LLMs have emerge as powerful tools for tackling summarization tasks. As LLMs are trained on massive datasets of text and code they have great exposure which allows them to learn complex relationships between words and grasp the nuances of language. Compared to traditional summarization methods, LLMs offer several advantages:

\begin{itemize}
\item \textbf{Superior Context Understanding:} Unlike simpler techniques that extract keywords, LLMs delve deeper, understanding the underlying meaning and context of the text \citep{SummaryLLM}. This enables them to capture the essence of the information more effectively.
	\item \textbf{Abstractive Summarization:} LLMs go beyond merely copying sentences. They can paraphrase, rephrase, and generate entirely new sentences, creating summaries that are fluent and human-like \citep{SummaryLLM}.
    \item \textbf{Adaptability:}LLMs can be fine-tuned to specific domains or summarization goals, making them versatile tools for various applications \citep{SummaryCompLLM}.
\end{itemize}

\begin{figure}[!b]
	\centering
	\includegraphics{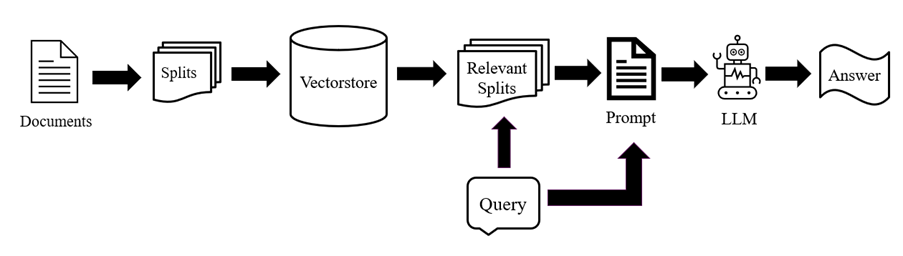}
	\caption{RAG Flow}
	\label{fig:fig1}
\end{figure}

\subsection{RAG (Retrieval-Augmented Generation)}
RAG (Retrieval-Augmented Generation) is an approach in natural language processing (NLP) that combines the strengths of retrieval-based systems and generative models \citep{RAGSurvey}. The core idea behind RAG is to enhance the generative model's responses by retrieving relevant information from a large corpus of documents. This approach helps in producing more accurate and informative answers, especially when dealing with complex queries or topics where the generative model alone might lack specific knowledge.

RAG based system involves the following process:
\begin{itemize}
	\item \textbf{Query Encoding:}  The input query is encoded into a dense vector representation using a pre-trained encoder, such as BERT or another transformer-based model.
	\item \textbf{Retrieval:} The encoded query is used to retrieve relevant documents or passages from a large corpus. This retrieval is often performed using similarity search techniques, such as FAISS  (Facebook AI Similarity Search)\citep{FAISS}, which quickly finds the most relevant documents based on vector similarity.
    \item \textbf{Document Encoding:} The retrieved documents are also encoded into dense vector representations.
    \item \textbf{Generation:} The encoded query and the retrieved document representations are fed into a generative model, such as a sequence-to-sequence transformer (e.g., T5 or GPT-3). The generative model uses this information to produce a more informed and contextually relevant response.
\end{itemize}
 
Figure \ref{fig:fig1}. shows the process flow for RAG. RAG based systems are commonly used for question answering, conversational agents, document summarization and knowledge base completion. RAG combines the strengths of information retrieval and large language models (LLMs). It leverages external knowledge bases to supplement the LLM's internal knowledge, overcoming limitations like:
\begin{itemize}
	\item \textbf{Knowledge Cutoff:}  LLMs have a limited knowledge base based on their training data. RAG bridges this gap by accessing external sources for real-time updates, enhancing the accuracy and reliability of its responses \citep{Nightfall}.
	\item \textbf{Hallucination Risks:}  LLMs can sometimes generate responses that are factually incorrect or irrelevant. RAG mitigates this risk by grounding the LLM's outputs in factual information retrieved from trusted sources \citep{Nightfall}.
    
\end{itemize}

This integration of retrieval and generation capabilities makes RAG particularly valuable for various cybersecurity tasks like:
\begin{itemize}
    \item\textbf{Threat Intelligence Analysis:} Automating the analysis of vast datasets allows for faster identification of patterns and anomalies indicative of potential threats. RAG can sift through threat intelligence feeds and security logs, summarizing findings and highlighting critical information for analysts \citep{threat}.
    \item \textbf{Incident Response:} RAG can assist in incident response by retrieving relevant knowledge base articles, security advisories, and remediation procedures based on the specific details of the incident. This expedites the response process and reduces downtime. One such example being Patch-RAG \citep{prag} which automates the process of generating step-by-step guides for addressing software vulnerabilities. 
    \item \textbf{Security Awareness Training:} RAG can generate personalized and informative training materials tailored to an audience's specific needs and knowledge level. It can also answer user queries regarding security policies and best practices, promoting a more security-conscious workforce.
\end{itemize}

By harnessing the power of RAG, we intend to provide a comprehensive and accurate summary of individual packets and their impact, which can be utilized to support analysts and in generating reports.

\section{Related Work}
\label{sec:related work}
Research papers highlight the growing interest in leveraging LLMs for various cybersecurity tasks. A recent study \citep{cyberSurvey} explores the potential of LLMs in tasks like:
\begin{itemize}
	\item \textbf{Vulnerability Detection:} LLMs can analyze vast amounts of code to identify patterns indicative of vulnerabilities.
	\item \textbf{Malware Analysis:} By understanding the language used in malware code and associated documentation, LLMs can aid in classification and analysis.
    \item \textbf{Network Intrusion Detection:} LLMs can analyze network traffic data to detect anomalies and potential intrusions. 
    \item \textbf{Phishing Detection:} The ability to understand natural language allows LLMs to identify suspicious emails with higher accuracy.
\end{itemize}
Another application of LLMs, more related to the problem addressed in this paper, is generating comprehensive and understandable reports based on network packet analysis. LLMs can learn the characteristics of malicious traffic data, detect anomalies in user-initiated behaviors, and describe the intent behind intrusions and abnormal behaviors. For instance, Fayyazi and Yang \citep{AmbiguousDesc} explored the use of LLMs to predict and differentiate between ATT\&CK tactics. Ali and Kostakos \citep{Huntgpt} implemented HuntGPT, an intrusion detection dashboard that combines a Random Forest classifier, XAI frameworks like SHAP and Lime, and GPT 3.5 Turbo to deliver threats in an easily understandable format. Furthermore, LLMs can provide security recommendations and response strategies for identified attack types \citep{PSA}.

LLMs are also employed in Cyber Threat Intelligence (CTI) reporting. SecureBERT, a cybersecurity language model developed by \citep{SecureBert}, is designed to understand the subtleties of cybersecurity text, enabling it to automate several critical cybersecurity tasks. \citep{ScriptKiddie} discussed how LLMs can understand threats, generate information about cybersecurity tools, and automate cyber campaigns.

\section{Problem Statement}
\label{sec:problem statement}
The Security Operations Center (SOC) analysts employ Intrusion Detection Systems (IDS) to scrutinize network traffic or system activity for indications of malicious activity or policy breaches. IDS alerts are generated when the system detects suspicious or anomalous activity that could signify a security threat, and these alerts are associated with specific packets. Analysts must comprehend why particular packets triggered an alert for efficient incident response. To gain this understanding, analysts need to be familiar with the specifications and protocols of control systems, which involves reading extensive technical documents and protocol specifications. Consequently, the training process is lengthy, and it takes time for new analysts to become proficient or for monitoring new environments to commence.

Not much literature exists on explanation of individual packets for packet analysis. With the emergence of large language models (LLMs), generating packet explanations and summarizations has become more feasible.

The simplest approach to generate packet explanations would be to use LLMs directly to interpret packet data and provide answers based on that understanding. However, this method relies heavily on the inherent capabilities of LLMs to comprehend packet files and their knowledge of networking protocols. Although most LLMs possess the necessary abilities to understand and describe packet data, they tend to provide more generic explanations in the absence of context.

An alternative option is to utilize existing retrieval-augmented generation (RAG) based systems, where the network standard serves as context, and the extracted packet information acts as a query. Relying solely on traditional RAG-based systems has demonstrated issues, such as retrieved data not providing sufficient context for an accurate response or the division of documents into smaller pieces leading to the loss of nuances and connections between different sections, resulting in incomplete content representations. In the case of packet data, retrieval based on current RAG methods tends to suffer more, as packet data focuses on keywords and contains less descriptive information in the query compared to typical RAG-based question-answering applications.

To tackle these challenges, we propose a RAG-based summarization system tailored to packet data. The solution addresses the above-mentioned issues by improving on the context retrieval method for RAG, resulting in more the accurate and relevant information being provided to LLM for summarization. 

\section{Proposed Solution}
\label{sec:Proposed Solution}
\begin{figure}[!b]
	\centering
	\includegraphics[scale=0.5]{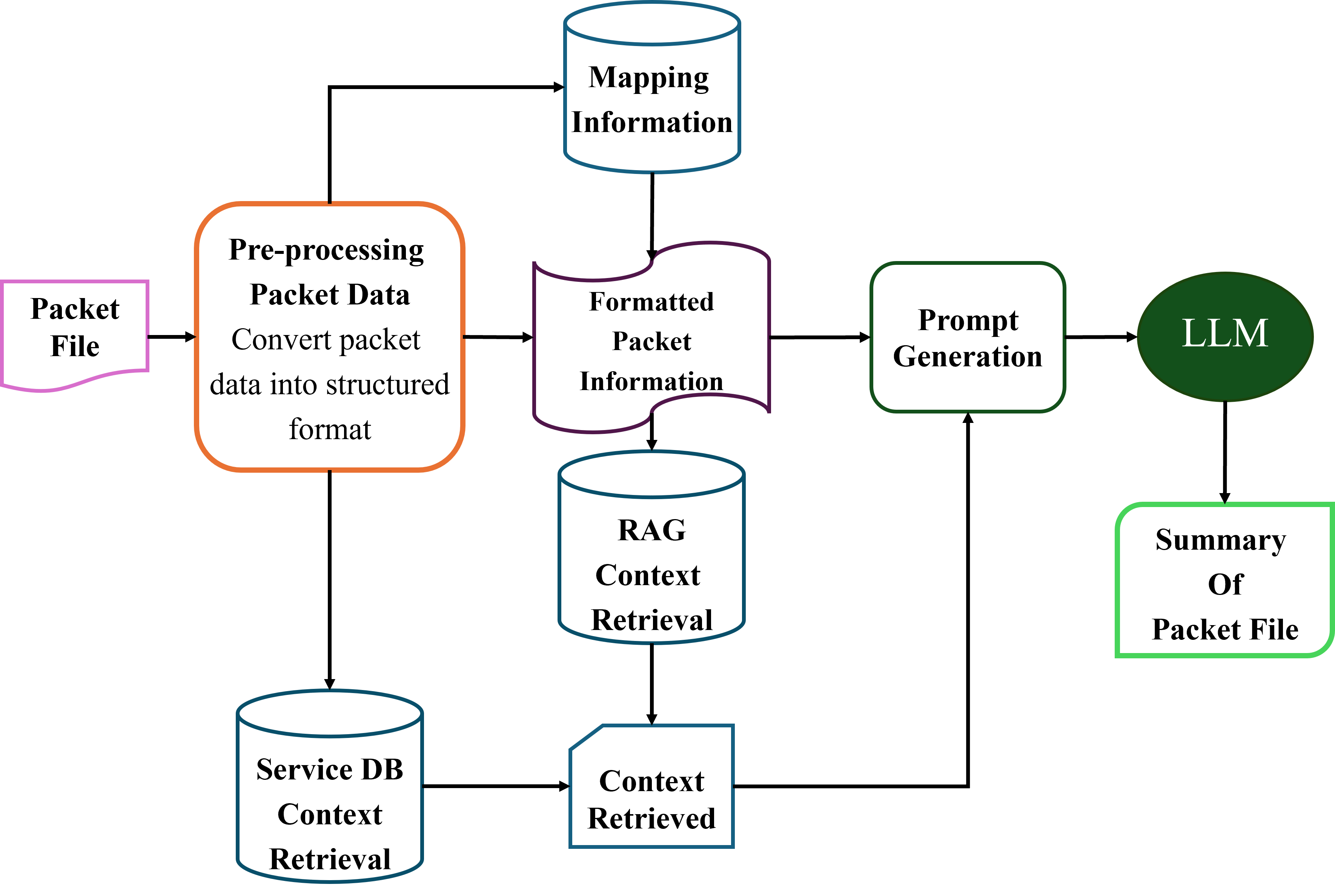}
	\caption{Proposed Solution Architecture}
	\label{fig:fig2}
\end{figure}
The overall architecture of our solution is illustrated in Figure \ref{fig:fig2}. The packet file is first pre-processed to extract useful information. This information is used to extract device information from the database. The packet information is also used for context retrieval using service database and RAG based context retrieval. Formatted packet information combined with context and device information is then inputted to the LLM which generates a packet file summary. Details of each stage are provided below:

\subsection{Packet Data Preprocessing}
Our solution leverages the structure of network packets to generate summaries. Although the structure of packets may vary depending on the protocol used, each packet typically consists of two primary parts: the header and the payload (or data). In the case of BACnet, a typical packet includes the following layers:
\begin{itemize}
	\item \textbf{Application Layer:} This layer manages BACnet services and encodes/decodes BACnet data. It contains Application Protocol Data Units (APDUs), which are the fundamental communication units in BACnet. APDUs carry information related to BACnet services, such as ReadProperty, WriteProperty, and WhoIs.
	\item \textbf{Network Layer:} This layer handles the routing and addressing of BACnet packets. It adds a Network Protocol Data Unit (NPDU) header to the APDU, which contains information about the source and destination addresses, network number, and message control information.	
 \item \textbf{Data Link Layer:}This layer is responsible for transmitting and receiving BACnet packets over the physical medium. It adds a Data Link Protocol Data Unit (DPDU) header and trailer to the NPDU, which contains information about the frame type, source and destination MAC addresses, and error detection.
\item \textbf{Physical Layer:}This layer deals with the physical transmission of BACnet packets over various communication media, such as Ethernet, ARCNET, MS/TP, and BACnet/IP. It defines the electrical and mechanical characteristics of the communication channel.
\end{itemize}
 
We extract information from the NDPU, DPDU and APDU layers. The NDPU and DPDU provides the following information: source and destination IP addresses, source and destination ports, and source and destination Organizationally Unique Identifier (OUI). Since the APDU layer contains the payload, all data from this layer is extracted. During preprocessing, all packets in a packet file are extracted, separated, and stored as individual units for further processing. 

\begin{figure}
	\centering
	\includegraphics[scale=0.5]{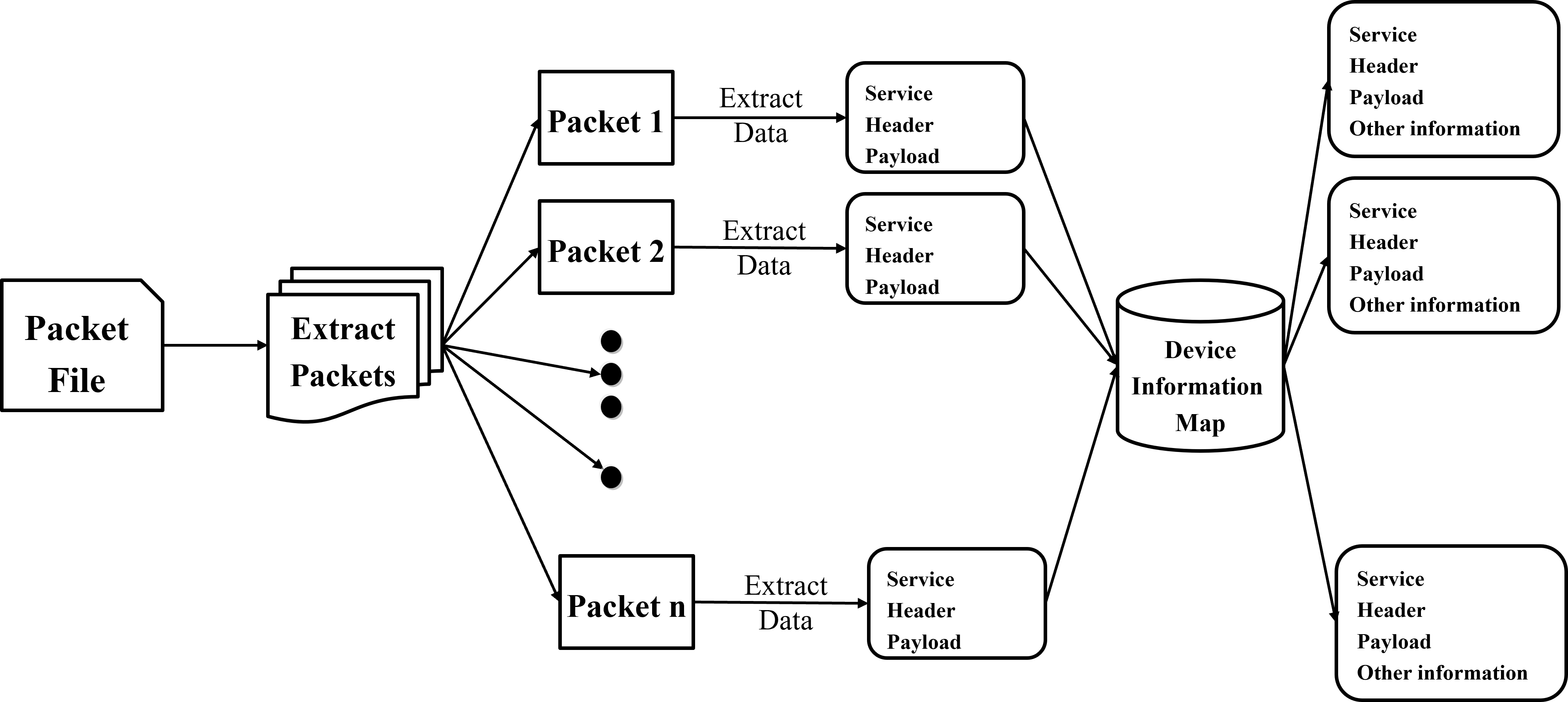}
	\caption{Mapping device information for all packets in packet file}
	\label{fig:fig3}
\end{figure}
\subsection{Mapping of Device Information}
We possess information about the devices communicating via the BACnet protocol for buildings monitored by our security services. We provide this information to the LLM for a better understanding of communicating devices in a packet file. This information is only used if available. We map this device information to packet file data as part of preprocessing. Figure \ref{fig:fig3} shows the breakdown of packet file into consistent packets and extraction of device information per packet. Adding this information to packet data provides insights to both LLM and SOC operators about which devices are affected by a particular packet. 
\subsection{Context Extraction}
\begin{figure}
	\centering
	\includegraphics[scale=0.5]{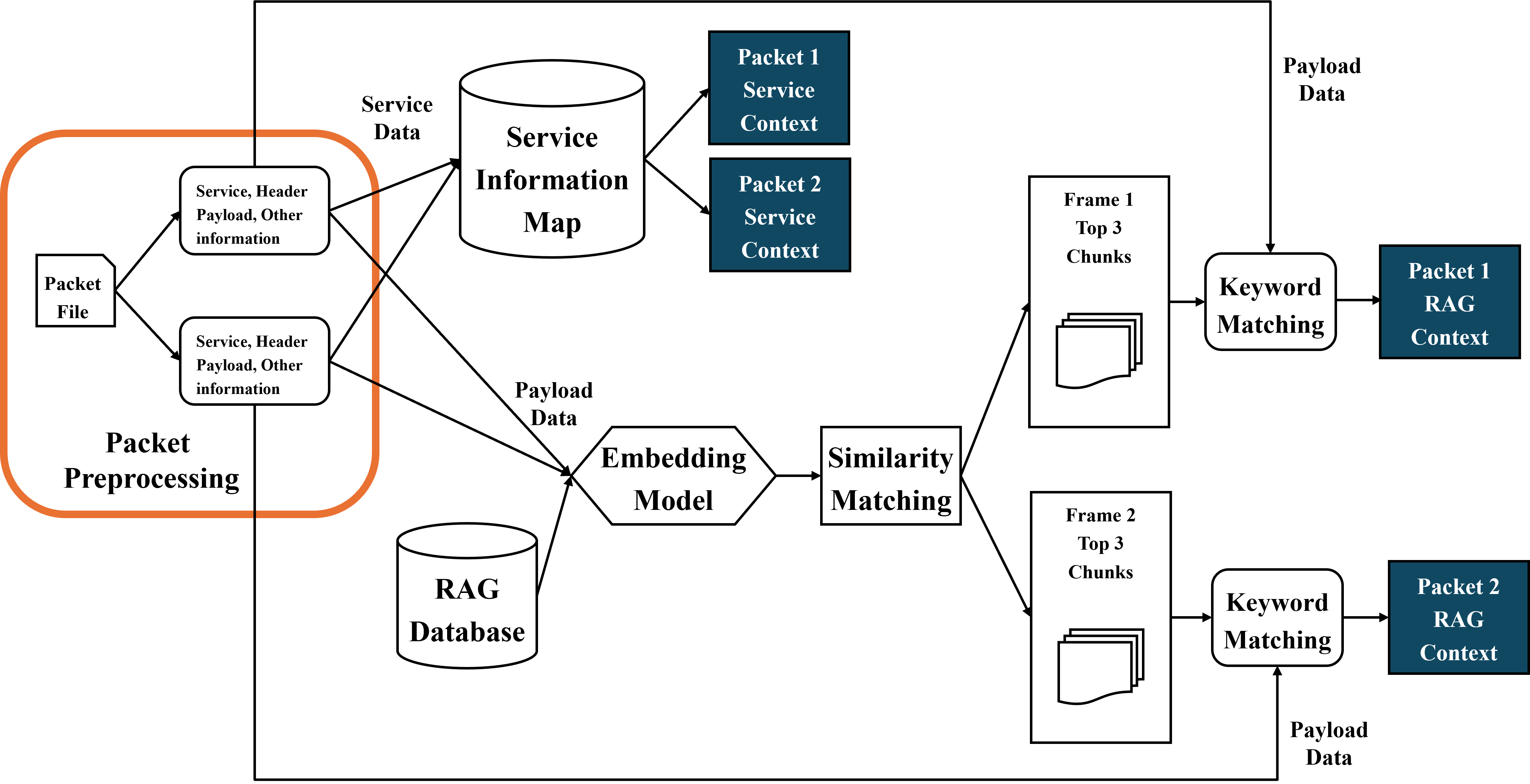}
	\caption{Context extraction for each packet of packet file with two packets}
	\label{fig:fig4}
\end{figure}

Context is extracted using two sources: service database and RAG based context retrieval. Figure \ref{fig:fig4} shows how information retrieval is carried out for a packet file with 2 packets. The details of context retrieval for each of the extraction method is in the following sections.  
\subsubsection{Service Database Context Extraction}
The APDU of a packet contains information about BACnet services. To ensure that LLM always receives service information as context for a given packet, we created a service database. This database contains summarizations of specific functions or actions performed by a service, mapped to the service name as the key. For each packet, the service information is extracted using this database and used as context for packet summarization.

\subsubsection{RAG-based Context Extraction}
To provide LLM with additional context beyond the operation performed by the packet, we created a RAG database for the remaining data. The data is divided into chunks using Llamaparse\footnote{\href{https://docs.llamaindex.ai/en/stable/llama_cloud/llama_parse/}{LLamaParse : https://docs.llamaindex.ai/en/stable/llama\_cloud/llama\_parse/}} and customized document-specific chunking. The chunks are stored using FAISS, a library for efficient similarity search and clustering of dense vectors.
For each packet, the APDU data is used as a query to extract the top 3 similar chunks from the RAG database. Keyword matching is performed between the APDU data and the extracted chunks to improve the accuracy of the extraction. The chunk with the most matching keywords and the highest rank is used as context in the LLM input. 

\subsection{Summary Generation}
The context extracted using the FAISS and service databases for each packet in the packet file is combined, with redundancies removed, and provided as context in the prompt. The processed packet data is included as part of the query to the LLM. Using this information, the LLM model generates a summary for the entire packet file.

\section{Experiments and Results}
\label{sec:Experiments and Results}
For embedding model, for RAG context retrieval, we used all-MiniLM-L12-v2 model. The LLM model used in our experiments was Mixtral 8X7B LLM model \footnote{\href{https://mistral.ai/news/mixtral-of-experts/}{Mixtral of experts : https://mistral.ai/news/mixtral-of-experts/}} installed on our local machine. To improve speed and reduce hardware requirements we made use of the 4-bit quantized version of the model.

For context we use the BACnet standard book: ASHRAE BACnet, A Data Communication Protocol for Building Automation and Control Networks\footnote{\href{https://store.accuristech.com/standards/ashrae-135-2020?product_id=2191852}{Standard 135-2020 -- BACnet -- A Data Communication Protocol for Building Automation and Control Networks (ANSI Approved)}}.

We performed the following experiments for comparison purposes:
\begin{enumerate}
  \item \textbf{Summarization generation without RAG (Method 1)} \\
  In summarization generation without RAG, we pass the processed packet content to the LLM and send the query to generate a summary for the packet, based on that information. As the packet data is converted into plain text after preprocessing the LLM provides information about the packet by explaining the available text.
  \item \textbf{Summarization generation using RAG without service database context(Method 2)} \\
  In summarization generation using only context retrieved using current RAG-based context retrieval methods without service database context, we pass the processed packet with context extracted using RAG based context retrieval. For chunking the standard, we make use of Llamaparse followed by recursive character text splitter. The chunks are saved in a FAISS database. For each packet in the packet file a chunk is retrieved from the database. All the extracted chunks are combined and provided as context along with the packet data to the LLM. The LLM then provides the summary based on the information.
  \item \textbf{Summarization generation using only service database context retrieval (Method 3)} \\
  In summarization generation using only service database, we pass the service information mapped to the service name in packet. The service information is extracted from each packet of the packet file as context and then combined with the packet data, is sent to LLM which then generates a response.
  \item \textbf{Summarization generation using the proposed solution (Method 4)}
\end{enumerate}

The first three methods are used for comparison purposes to the proposed solution.

\subsection{Evaluation}
The task dealt in this paper is a controlled task, limited to summary generation for BACnet packets. So, we do not possess a standard reference dataset for evaluation. For evaluation purposes we make use of 19 BACnet packets with IDS alerts(which can be used in future evaluations when IDS information is incorporated in summary). 

For tasks like extractive summarization, where full source sentences are selected to appear in the summaries,
simple n-gram overlap metrics against the “gold”
summaries like ROUGE\citep{Rouge} or BLEU \citep{Bleu} tend to work well as the
correct answer space is limited \citep{Eval}. In our task multiple summaries might be equally good so instead of creating "gold" summaries for our evaluation we proceed with human evaluation. Specifically we involve people with knowledge about BACnet protocol to evaluate the summaries based on two criteria: (1) Control Accuracy (CA) \citep{HEval}: whether the summary contains accurate information about the intent of the packet and does not contain information which might mislead the analyst. (2) Control Informativeness(CI): the extent to which the summary can provide useful information about the packet. Results of evaluation are shown in Table \ref{tab:table}.

\FloatBarrier
\begin{table}[H]	
	\centering
	\begin{tabular}{lll}
		\toprule		
		Method     & CI     & CA \\
		\midrule
		Method 1 & 3.18  & 4.86     \\
		Method 2     & 3.26 & 4.6      \\
		Method 3     & 3.23  & \textbf{4.89}  \\
            Method 4     & \textbf{3.63}  & \textbf{4.89}  \\
		\bottomrule \\
	\end{tabular}
 \caption{Human evaluation on packet summary relevance experiments (scale 1-5, higher is better) across all methods. Control accuracy (CA) and control informativeness (CI) are reported. }
	\label{tab:table}
\end{table}

\subsection{Result Analysis}
 From Table \ref{tab:table} it can be gleaned that the proposed solution outperforms the other 3 methods in terms of informativeness while maintaining a high accuracy in the information provided.

In method 1 where the LLM is not provided any context information, the method performs poorly, for the CI metric, compared to other methods reasonably. In absence of any context the model hallucinates or provide any wrong information due to irrelevant context, rarely, resulting in a good accuracy of summary generated. In comparison to the proposed method(method 4), method 1 obviously lags in terms of informativeness and slight difference in accuracy due to absence of relevant contextual information.

In method 2 the context is retrieved entirely using the SOTA context retrieval methods. However, the context extracted is not always pertinent to the packet data leading to a poor accuracy of the method compared to others. The main downside observed in this method was irrelevant context extraction resulting in misleading information in the summary. 

Method 3 focused on providing appropriate context for the packet. This resulted in higher accuracy of the information provided. The information though was not all encompassing as it was limited just to service part of the payload. This caused the slight difference in performances, in terms of informativeness, compared to method 2 which provides context for the entire payload. 

In method 4, our proposed solution we tried to reduce the drawbacks of the previous methods while retaining their benefits. We improved upon context extraction from method 2 to make context more relevant and used context extraction from method 3 to ensure accurate information of packet intention is available to the LLM. This resulted in summaries with better informativeness compared to other solutions while maintaining a high accuracy in the information provided.

\section{Conclusion}
\label{sec:Conclusion}
In this paper, we discussed methods for generating packet data summaries to extract relevant information from network packets. We proposed a Retrieval-Augmented Generation (RAG) approach that combines existing context retrieval techniques with mapping-based context retrieval to produce comprehensive summaries. This solution can assist SOC operators in quickly understanding packet data, help train new analysts, and provide clear explanations to non-technical stakeholders.

The proposed solution however suffers from the drawback of extracting large amount of context, if a packet file has lot of packets, which can result in the LLM running out of context window. In the future, we aim to address current limitations, such as handling packet files with a large number of packets and extracting more focused context for our summaries.  We also plan to expand our solution beyond BACnet packets to other ICS networks, incorporate additional information, such as Intrusion Detection System (IDS) alerts and improve our evaluation data accordingly.


\bibliographystyle{unsrtnat}
\bibliography{references}

\clearpage
\appendix
\section{Output Examples}
\label{sec:Examples}
In this section, we show examples of a few summaries generated by our proposed solution. For privacy reasons some information has been omitted.

\FloatBarrier
\begin{table}[H]	
	\centering
	\begin{tabular}{p{4cm}p{12cm}}	
		\midrule
		Formatted Packet File & \parbox{12cm}{\begin{flushleft}
  Packet:\\
Frame 1  \\
  adpu\_type :Confirmed-REQ \\
  adpu : \\   Service Choice: writePropertyMultiple (16)\\
    ObjectIdentifier: analog-output, 28, Name :AC-100 Server room temperature setting, Type :AirConditioning \\
    Property Identifier: present-value (85) \\
    Present Value (real): 100 \\
    Priority: (Unsigned) 1
    \end{flushleft}}
     \\
     \hline
Summary Generated & \parbox{12cm}{\begin{flushleft}
The given packet is a BACnet (Building Automation Control network) packet, specifically a Confirmed Request (Confirmed-REQ) message.\newline 

The packet contains a single Application Layer Service called "writePropertyMultiple" (16), which is used to modify multiple properties of a         BACnet object. In this case, it is being used to change the "present-value" property  of a specific Analog Output object. \newline

The Analog Output object is identified by the ObjectIdentifier field, which includes the object type ("analog-output"), instance number (28), and name ("AC-100 Server room temperature setting"). The object type belongs to the category "AirConditioning".
\newline

The Property Identifier field specifies that the "present-value" property of the Analog Output object is being modified. The Present Value field contains the new value     for the property, which is set to 100.
\newline

Finally, the Priority field indicates the priority level assigned to this write operation, with a value of 1 being the highest priority.
\newline

As a result, the air conditioning system associated with the Analog Output object will adjust its server room temperature setting to 100 units based on the priority level 1.
 \end{flushleft}
}
\\
		\hline
	\end{tabular}
\end{table}

\begin{table}	
	\centering
	\begin{tabular}{p{4cm}p{12cm}}	
		\midrule
		Formatted Packet File & \parbox{12cm}{\begin{flushleft}
  Packet:\\
Frame 1  \\
  adpu\_type :Confirmed-REQ\\
  adpu :  \\  Service Choice: writeProperty (15)\\
    ObjectIdentifier: device, 126\\
    Property Identifier: object-list (76)\\
    object-list: NULL\newline
    
Frame 2 \\ 
  adpu\_type :Error\\
  adpu : \\       Service Choice: writeProperty (15)\\
    Error Class: property (2)\\
    Error Code: write-access-denied (40)
    \end{flushleft}}
     \\
     \hline
Summary Generated & \parbox{12cm}{\begin{flushleft}
The given packet contains two frames. The first frame is a Confirmed Request (Confirmed-REQ) to perform the "writeProperty" service, which is used to change the value of a specific property of a BACnet object. In this case, the object identifier is "device" with instance number 126, and the property identifier is "object-list" with identifier 76. The "object-list" property contains a list of object identifiers that are associated with the device. The value of "object-list" in this frame is NULL, meaning that the client is asking the server to clear the existing list of object identifiers associated with the device.\newline

However, the second frame is an Error response, indicating that the write-access to the "object-list" property was denied. The Error Class is "property", and the Error Code is "write-access-denied". This means that the server did not allow the client to modify the "object-list" property because it is defined as inaccessible through the BACnet protocol write services. Therefore, the client's request to clear the existing list of object identifiers was rejected.
\newline

In simpler terms, the client tried to update the list of objects associated with the device but the server denied the request because it doesn't allow changes to that particular property.

 \end{flushleft}
}
\\
		\hline
	\end{tabular}
\end{table}

\end{document}